**Title: Reproduction number of SARS-CoV-2 Omicron variants, China, December 2022-January 2023**

**Running Title.** Reproduction number of SARS-CoV-2 Omicron variants in China

**Authors:** Yuan Bai[1,2], Zengyang Shao[2], Xiao Zhang[2], Ruohan Chen[2], Lin Wang[3], Sheikh Taslim Ali[1,2], Tianmu Chen[4], Eric H. Y. Lau[1], Dong-Yan Jin[5], Zhanwei Du[1,2*]

**Affiliations:**

[1] WHO Collaborating Centre for Infectious Disease Epidemiology and Control, School of Public Health, Li Ka Shing Faculty of Medicine, The University of Hong Kong, Hong Kong Special Administrative Region, China

[2] Laboratory of Data Discovery for Health, Hong Kong Science and Technology Park, Hong Kong Special Administrative Region, China

[3] Department of Genetics, University of Cambridge, Cambridge CB2 3EH, UK

[4] State Key Laboratory of Molecular Vaccinology and Molecular Diagnostics, School of Public Health, Xiamen University, Xiamen, China

[5] School of Biomedical Sciences, Li Ka Shing Faculty of Medicine, The University of Hong Kong, Hong Kong Special Administrative Region, China

YB, ZS, XZ and RC contributed equally to this manuscript.

Corresponding authors: Zhanwei Du (zwdu@hku.hk)

**Highlight**

China adjusted the zero-COVID strategy in late 2022, triggering an unprecedented Omicron wave. We estimated the time-varying reproduction numbers of 32 provincial-level administrative divisions from December 2022 to January 2023. We found that the pooled estimate of initial reproduction numbers is 4.74 (95% CI: 4.41, 5.07).

**Main Text**

China had successfully suppressed multiple waves of SARS-CoV-2 epidemics using the "dynamic zero-COVID" strategy before November 2022 [1]. Due to the substantially reduced pathogenicity of new SARS-CoV-2 Omicron variants (e.g., BF.7 and BA.5.2) in populations with higher vaccine coverage, and enormous socioeconomic costs incurred by the dynamic zero-COVID strategy, China began to adjust response strategies from 11 November 2022 (including restricting testing coverage, shortening quarantine periods for inbound travelers and suspending secondary contacts tracing) [2]. Starting from 7 December 2022 [3], the further relaxation of control measures (including the prohibition of regional mass testing and the implementation of home isolation or quarantine) triggered an unprecedentedly large Omicron wave in China, leading to a sharp increase in fatalities. The abrupt change of COVID control measures and the resulting surge of SARS-CoV-2 infection, hospitalization and death in China have raised great international concerns [4].

The time-varying reproduction number ($R_t$), a measure of instantaneous transmissibility of an epidemic, is defined as the average number of secondary infections caused by a typical primary case at time $t$, after accounting for the population immunity and the impact of control measures. Reliable estimates of $R_t$ are essential to understand how the transmissibility of COVID-19 changes after the relaxation of the public health and social measures (PHSMs). As the population-wide testing of SARS-CoV-2 was abandoned in mainland China from 8 January 2023, surveillance data about the daily number of new infections were lacking since then. To fill this gap, on 25 January 2023 the Chinese Center for Disease Control and Prevention (CDC) published a unique surveillance data estimating the daily hospitalization number of COVID-19 patients in hospitals from December 2022 to January 2023 in China [5]. Informed by these data, we evaluate $R_t$ for the 32 provincial-level administrative divisions in the study period, together with the number of daily new cases to nowcast the epidemic growth of COVID-19 Omicron variant in China from December 2022 to January 2023.

We evaluated the time-varying reproduction numbers ($R_t$) of the COVID-19 Omicron outbreaks in mainland China at the 32 provincial-level administrative divisions (**Figure 1**, **Supplementary Methods**). The prevalence of hospitalizations peaked in early January 2023, with variations across administrative divisions. The peak prevalence of hospitalization (per 100,000 population) ranged from 22 in Tianjin to 390 in Ningxia.

Informed by the hospitalization data, we estimate the daily $R_t$ at the provincial-level administrative divisions. We found that the initial $R_t$ among the 32 study divisions ranges from 3.10 (95% confidence interval [CI]: 2.71, 3.49) in Xinjiang Production and Construction Corps (XPCC) to 7.55 (95% CI: 4.48, 10.63) in Jiangsu province. Of these, the reported subvariants in the surveillance system mainly include BA.5.2 (70.2%), predominant in southern regions, and BF.7 (28.3%), predominant in northern regions [5].

The meta-regression analysis was conducted based on the reported initial $R_t$ estimates for the Omicron variant, which allowed us to explore the potential association between the location and $R_t$. High heterogeneity was reported among divisions ($I^2 = 0.71$, $p < 0.01$) (**Figure S1**). The pooled estimate of initial $R_t$ was 4.74 (95% CI: 4.41, 5.07) in the 32 provincial-level administrative divisions.

From August 2021, China implemented a "dynamic zero" COVID strategy until November 2022. This was followed by a rapid transition between partial relaxation on 11 November, 2022, and complete opening-up on 7 December, 2022. Considering the low vaccination rates among those over 80 years of age and the low population immunity acquired through previous infections, China's elimination of zero-clearance could trigger a wave of Omicron infections under a high transmission scenario that might result in over 90% of the population being infected as of 19 January 2023 [5]. We found high heterogeneity of initial reproduction numbers across divisions in mainland China and the pooled estimate is 4.74 (95% CI: 4.41, 5.07).

China has its Spring Festival on 22 January 2023, for which residents and visitors normally made several billion trips throughout China to celebrate the Lunar New Year, starting on 7 January in 2023. However, given the early starting of nationwide loosening, the epidemic impact of the Festival travel might have fueled the transmission in provinces (e.g., Ningxia) (**Figure 1.A**) .

The basic reproduction number of the ancestral variant was estimated to be 3.15 (95% CI: 3.04, 3.26) during January 2020 in Chinese provinces [6], and the initial reproduction number of the Omicron outbreak was estimated to be 3.44 (95% credible interval [CrI]: 2.82, 4.14) in Beijing in mid November 2022 based on Weibo and WeChat online polls, when the Omicron wave started in Beijing several weeks earlier than in other parts of mainland China [7]. Our pooled estimate in this study is higher than those of earlier COVID-19 waves, denoting the unprecedented COVID-19 outbreaks in mainland China given the low population immunity in December 2022.

Globally, in 2021 and 2022, the effective reproduction numbers of the Omicron outbreaks were estimated for South Africa, Denmark, China, England, and India [8]. The mean estimates ranged from 2.43 to 5.11, with the pooled estimate being 4.20 (95% CI: 2.05, 6.35) [8]. These estimates are close to our pooled estimates in mainland China, suggesting similar transmission efficiency across countries in the initial Omicron outbreaks.

Although the COVID-19 outbreak almost ended in China in late January 2023, the World Health Organization (WHO) claimed that the COVID-19 pandemic continued to constitute a public health emergency of international concern on 27 January 2023 [9]. We would provide an overview of key assumptions in this study. First, our model might be demographically biased by the surveillance system of China CDC [5], which is based on sentinel hospitals. Criteria of hospital admissions and discharge might have changed when the hospitals reached their capacity, which might have biased the reproduction number estimates. Further, considerable uncertainty remained regarding the lag between hospitalizations and infections. We appeal to the researchers to be careful about their conclusions when using this time-varying reproduction number. We expect that estimates for the time-varying reproduction number will improve as more granular epidemiologic data become available. However, our key qualitative insights are likely robust to these uncertainties.

Our study could inform the relevant governing authorities to make the necessary choices regarding the prioritization of conventional and emergency health systems in the next pandemic. Given the transition from the zero-COVID policy to the major relaxation in a short period, China would be expected to prepare in advance by training millions of healthcare workers to care for mild COVID-19 cases and stockpiling antivirals. Compared with earlier COVID-19 waves in China, most people were infected in this outbreak in a short period, and the country has likely leaped from a containment phase to a recovery phase through the abrupt relaxation of control measures [4]. The released surveillance data from Chinese CDC sentinel household surveillance system [5] should provide another opportunity to learn more about the resulting surge and reduce international concerns, after China disbanded its national case reporting system [10].

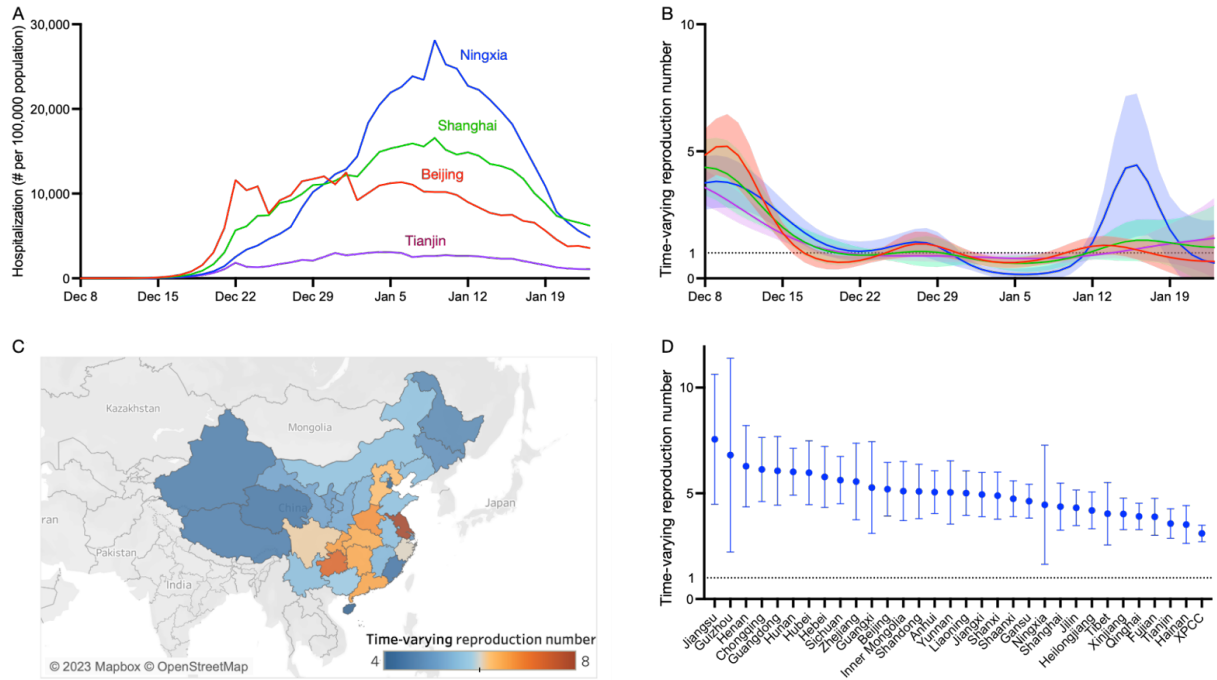

**Figure 1. Estimated time-varying reproduction number of regions in mainland China between 8 December 2022 to 23 January 2023.** (A) Hospital prevalence (# per 100,000 population) in Beijing, Shanghai, Tianjin, and Ningxia between 8 December 2022 to 23 January 2023. (B, C, D) Time-varying reproduction number ($R_t$) in the provincial-level administrative divisions, estimated from hospitalization data. (B) Lines and shaded areas indicate the mean and 95% CI of $R_t$ over time in Beijing (red) and Shanghai (green). (C) Mean estimates of initial $R_t$ presented by provincial-level administrative divisions. (D) Mean and 95% CI of initial $R_t$ estimates in the study period presented by divisions.

## Author Contributions

YB, ZS, XZ, RC and ZD: conceived the study, designed statistical and modeling methods, conducted analyses, interpreted results, wrote and revised the manuscript; EHYL, DJ, TC, STA, and LW,: interpreted results and revised the manuscript.

## Funding

This work was supported by grants from the AIR@InnoHK Programme from Innovation and Technology Commission of the Government of the Hong Kong Special Administrative Region. The funders had no role in the design and conduct of the study; collection, management, analysis, and interpretation of the data; preparation, review, or approval of the manuscript; or decision to submit the manuscript for publication.

## Competing interests

The authors report no potential conflicts of interest.

## Data availability

All data are collected from open source with detailed description in Supplementary Method.

## Code availability

Code used for data analysis is freely available upon request.

# Supplement

**METHODS**

**Time-varying Reproduction Number Estimation**

The daily number of COVID-19 patients in hospitals ($H_t$) were extracted for each of the 32 provincial-level administrative divisions from the Chinese CDC surveillance dataset between December 2022 to January 2023 [5].

We first estimated the daily number of new hospital admissions ($h_t$), as given by

$$h_t = H_t - H_{t-1} + \sum_{k=1}^{t-1} h_k P_h(t-k) \qquad (1)$$

where $P_h(t-k)$ describes the duration that a COVID-19 patient newly hospitalized at day $k$ stayed in the hospital for treatment until discharged or died at day $t$, and follows a gamma distribution with shape=22.10, scale=1.33 [11].

We then estimated the daily number of cases from the time series of $h_t$, following a gamma distribution (with mean=6.93, sd=3.25) that describes the duration from infection to hospitalization [11] using the backprojNP deconvolution algorithm.

At last, $R_t$ is calculated with respect to the daily cases using the EpiNow2 package, by assuming a gamma distribution (mean=2.76 days, sd=0.53 days) of the generation time (the time interval between the infection of a primary case and one of its secondary cases) [12] and a gamma distribution (mean=4.21, sd=1.41) of the incubation time (the time interval between the onsets of symptoms of an index case and one of its secondary cases) [12]. We estimate the initial $R_t$ as the largest value in the first week of the study period from 8 December to 14 December 2022).

**Statistical Analysis**

To assess heterogeneity between locations, we used the $I^2$ index and grouped results into three categories: low heterogeneity for $I^2$ <25%, medium heterogeneity for $I^2$=25-75%, and high heterogeneity for $I^2$ >75%. Given the high $I^2$ value calculated in our results and the significant Cochran Q test, we chose to use a common-effects model for the meta-analysis in this study. Finally, we conducted meta-regression analysis using a mixed-effects model to quantify the association between the division and the estimate of the reproduction number. All analyses were performed using R version 4.1.1.

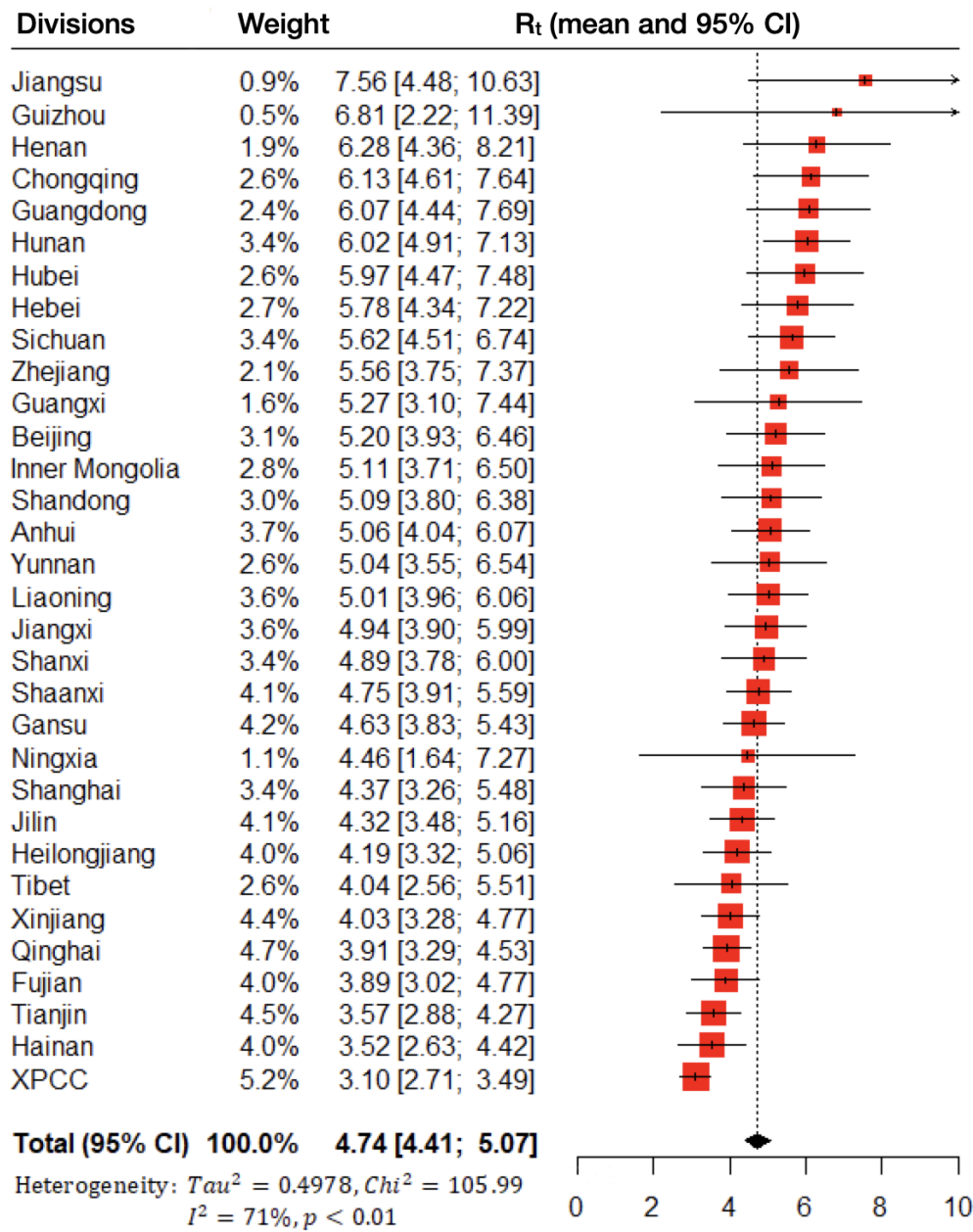

**Figure S1. Forest plot of the estimated reproduction numbers of Omicron variants from 32 provincial-level administrative divisions.** CI, confidence interval. XPCC, Xinjiang Production and Construction Corps.